\documentclass{article}

\usepackage{PRIMEarxiv}
\usepackage{amssymb}
\usepackage[T1]{fontenc}
\usepackage{graphicx}
\usepackage{color}
\usepackage{color, soul}
\usepackage{microtype}
\soulregister\cite7
\soulregister\ref7
\soulregister\pageref7
\usepackage{multirow}
\usepackage{comment}
\usepackage{amssymb}
\usepackage{pifont}
\usepackage{amsmath} 
\usepackage{algorithm2e}
\usepackage{booktabs}
\usepackage{subcaption}
\usepackage{multirow}
\usepackage{dsfont}
\usepackage{url}

\usepackage{tikz}
\usetikzlibrary{arrows.meta, positioning, shapes, decorations.pathreplacing, calc, 3d}
\usepackage{pgfplots}
\pgfplotsset{compat=1.17}

\title{CVE-TTP KG: Knowledge Graph Linking Software Vulnerabilities to Attack Behaviors}

\author{
Swati Yadav\\
Department of Computer Science \& Engineering\\
Swami Keshvanand Institute of Technology, 
Management \& Gramothan, Jaipur\\ 
\texttt{swati.yadav@skit.ac.in}\\
\And
Dincy R. Arikkat\\
Department of Computer Science and Engineering\\
Christ College of Engineering Thrissur, Kerala\\
\texttt{dincyrarikkat.cse@cce.edu.in}\\
\And
Basant Agarwal\\
Department of Computer Science \& Engineering\\  
Central University of Rajasthan, Ajmer\\
\texttt{basant@curaj.ac.in} \\
\And
Serena Nicolazzo \\
Department of Electrical, Computer and Biomedical Engineering,\\
University of Pavia, Italy\\
\texttt{serena.nicolazzo@unipv.it} \\
\And
Antonino Nocera\\
Department of Electrical, Computer and Biomedical Engineering,\\
University of Pavia, Italy\\
\texttt{antonino.nocera@unipv.it} \\
\And
Vinod P. \\
Department of Computer Applications, \\
Cochin University of Science \\
and Technology, India \\
\texttt{vinod.p@cusat.ac.in} \\
}

\begin{document}

\maketitle
\date{July 2026}

\begin{abstract}
In the evolving threat landscape, adversaries exploit software vulnerabilities to launch sophisticated attacks, challenging traditional defenses. Although databases like CVE and NVD provide detailed technical information, they often lack links to attacker behaviors such as tactics and techniques, limiting effective threat interpretation and response. This work bridges this gap by connecting vulnerabilities with behavioral patterns from the MITRE ATT\&CK framework. We construct a CVE–TTP Knowledge Graph that links CVEs to tactics and techniques using classification and relation extraction. Transformer-based models are developed for behavior identification, with CySecBERT achieving macro F1-scores of $87.71\%$ (techniques) and $96.16\%$ (tactics). Also, we created an annotated dataset with $24,820$ entities and $43,608$ relations for entity and relation extraction. The pipeline-based approach achieves macro F1-scores of 0.86 (entity extraction) and 0.99 (relation extraction), while a span-based joint model achieves 0.78. These outputs are integrated into a Neo4j-based Cyber Threat Knowledge Graph, enabling structured visualization of vulnerabilities.
\end{abstract}

\keywords{Cyber Threat Intelligence, Knowledge Graph, MITRE ATT\&CK, Vulnerability Analysis, Joint Entity Relation Extraction.}

\section{Introduction}

In the contemporary cybersecurity landscape, software vulnerabilities have become strategic assets that are actively traded, weaponized, and integrated into coordinated, economically motivated cyber campaigns by threat actors \cite{di2022software}. Each disclosed vulnerability represents a potential entry point into critical systems and is often exploited as part of broader attack operations \cite{makrakis2021vulnerabilities}. To standardize the reporting of such flaws, repositories like the Common Vulnerabilities and Exposures (CVE) \footnote{\url{https://cve.mitre.org/}} and the National Vulnerability Database (NVD)\footnote{\url{https://nvd.nist.gov/}} have been developed. Although these repositories offer structured identifiers and metadata for known vulnerabilities, they frequently lack operational context, such as attacker objectives, techniques, and exploitation paths needed to understand how vulnerabilities are actually used in real-world threat scenarios. Without this context, security teams are left with large volumes of vulnerability data but limited insight into which flaws are most likely to be exploited and how.

Traditional defensive strategies focus on patching and mitigation \cite{dissanayake2022software}. However, with the increasing sophistication of attacks, technical fixes alone are insufficient. To achieve cyber resilience, defenders must understand the intent behind adversary behavior and how specific vulnerabilities are linked to strategic goals. Frameworks like MITRE ATT\&CK \footnote{\url{https://attack.mitre.org/}} provide structured knowledge of Tactics, Techniques, and Procedures (TTPs), where \textit{tactics} represent the attacker’s goals or objectives (the \textit{what}), and \textit{techniques} describe the general methods used to achieve those goals (the \textit{how}). Integrating behavioral intelligence with vulnerability data is essential for prioritizing threats and anticipating attack patterns based on adversary intent and capability. The urgency of this need is magnified by the rapid growth in vulnerability disclosures. In 2024 alone, over 40,000 new CVEs were reported, marking a $39\%$ increase from the previous year\footnote{\url{https://socradar.io/top-10-exploited-vulnerabilities-of-2024/}}. This explosion of data makes manual analysis difficult and necessitates automated solutions that can process, correlate, and contextualize vulnerabilities at scale\footnote{\url{https://www.bitsight.com/blog/2025-predictions-for-cve-vulnerabilities}}.

To support contextualized vulnerability analysis, we propose a CVE-TTP Knowledge Graph (CVE-TTP KG) that links software vulnerabilities to attacker behaviors, specifically, the tactics and techniques. Our system first identifies relevant ATT\&CK tactics and techniques associated with the vulnerabilities using security-focused transformed models. Furthermore, the framework supports the extraction of key vulnerability entities and their relationships in the form of semantic triples: \textit{\textless entity1, relation, entity2 \textgreater}. These triples serve as the foundational input for constructing the CVE-TTP KG. 

This KG addresses a critical gap in current vulnerability analysis frameworks, which often treat vulnerabilities as isolated records lacking behavioral context. By explicitly modeling the connections between CVEs and adversary behaviors, our approach enables a deeper understanding of exploitation patterns, threat objectives, and attack chains. The CVE-TTP KG not only supports visualization and situational awareness for human analysts but also facilitates machine reasoning for automated threat detection, risk prioritization, and decision-making. 
The primary contributions of this work are outlined as follows:

\begin{itemize}
    \item   We introduce novel datasets for multi-label classification of ATT\&CK techniques and tactics, and a manually annotated dataset of vulnerability entities with semantic relations to support knowledge extraction and graph-based security analysis.

    \item 
    We implement a transformer-based classification model to automatically map software vulnerabilities to relevant ATT\&CK tactics and techniques. 

    \item   We develop models for extracting entities and their semantic relationships from vulnerability descriptions using both pipeline and joint learning approaches. 
    We conduct a comparative evaluation of these two strategies to assess their effectiveness in capturing vulnerability-specific relations.

    \item We construct a comprehensive KG that links CVEs to associated attack behaviors along with other relevant vulnerability information. The resulting graph offers a structured and interpretable representation of threat intelligence, enabling more effective analysis of vulnerability exploitation across adversarial campaigns.

\end{itemize}

The rest of the paper is organized as follows: Section \ref{sec:related_work} reviews related work. Section \ref{sec:architecture} details the proposed architecture and methodology. Section \ref{sec:experiment_results} presents the experimental results along with a comprehensive analysis. Finally, Section \ref{sec:conclusion} concludes the study and outlines directions for future research.

\section{Related Work}
\label{sec:related_work}
This section reviews relevant previous work under two main themes: (1) CVE-to-TTP classification, and (2) Threat KG construction for vulnerability-centric threat modeling.
\subsection{CVE-TTP Classification}
Recent years have seen growing efforts to map CVEs to MITRE tactics and techniques. Simonetto et al.~\cite{simonetto2024comprehensive} proposed a rule-based pipeline that leverages structured data such as CWE and semantic similarity to align CVEs with techniques. However, their approach relied on classical machine learning and lacked automation and contextual reasoning using modern NLP techniques. Kuppa et al.~\cite{kuppa2021linking} introduced a deep learning-based model that uses pseudo-labeled data to train on semantic matches between CVE text and ATT\&CK technique descriptions. Although effective in technique prediction, their work did not incorporate tactic-level classification or structured threat modeling.

Recent advances in using Large Language Models (LLMs) have opened new avenues. Hu et al.~\cite{hu2024llm} presented LLM-TIKG, a prompting-based framework that enhanced annotation and classification in cybersecurity tasks. However, their approach required manual intervention and was not tailored to CVE descriptions. In contrast, our work enables end-to-end, automated multi-label classification of both tactics and techniques and releases annotated datasets for benchmarking. Zhang et al.~\cite{zhang2024unittp} proposed UniTTP, a unified framework for the extraction of TTP across various cybersecurity inputs using prompt-based encoder-decoder models. While their system handles joint reasoning across sources such as Indicators of Compromises (IoCs) and reports, it lacks a specific focus on CVE-driven threat modeling. 
\subsection{Threat Knowledge Graph Construction}
Constructing effective threat KGs depends on accurately identifying relevant entities and their relationships within cybersecurity data. 
Several works have constructed cybersecurity KGs from open-source or structured threat intelligence. Xiao et al.~\cite{xiao2019embedding} explored embedding-based models to detect relationships among software entities, yet did not target ATT\&CK TTP semantics or vulnerability-to-technique links.  Shi et al.~\cite{shi2024uncovering} and Falcarin et al.~\cite{falcarin2024building} integrated heterogeneous data sources into graph-based representations but lacked direct extraction from CVEs or a focus on MITRE TTPs.

Gao et al.~\cite{gao2024threatkgaipoweredautomatedopensource} introduced ThreatKG, an automated KG generation system from CTI reports using semantic enrichment. While rich in contextual information, their model does not specifically connect CVEs to ATT\&CK frameworks. Similarly, Hu et al.~\cite{hu2024llm} built AttackKG using LLM-driven joint NER and RE on CTI reports, with limited attention to CVE alignment. Other notable efforts include CSKG4APT by Ren et al.~\cite{ren2022cskg4apt}, focused on attribution of APT campaigns via integration of malware and TTPs, and OSTIS by Arikkat et al.~\cite{arikkat2024ostis}, which constructed organizational threat KGs. Though both frameworks support advanced threat modeling, they do not explicitly extract or link CVE descriptions to ATT\&CK techniques. Zhang et al.~\cite{zhang2024attackgboostingattackknowledgegraph} proposed AttacKG+, combining LLMs with automated CTI parsing to generate ATT\&CK-aligned graphs. Their methodology shares similarities with ours in leveraging LLMs, though they primarily operate on threat reports. Our contribution bridges this gap by applying joint entity-relation extraction directly on CVEs and constructing ATT\&CK-grounded graphs for structured visualization of adversarial behavior.

Existing approaches primarily focus either on CVE-to-TTP mapping or KG construction, with limited efforts toward integrating both in a unified framework. They also lack joint NER and relation extraction tailored to CVE texts and provide limited support for ATT\&CK-based visualization of adversarial behavior. To address these gaps, we propose an integrated framework that performs multi-label classification of ATT\&CK tactics and techniques from CVE descriptions, applies joint entity and relation extraction for enriched threat intelligence, and constructs a vulnerability-centric KG for structured representation and visualization of adversarial behavior.

\section{Methodology}
\label{sec:architecture}
\begin{figure}[ht]
    \centering
    \includegraphics[width=0.8\textwidth]{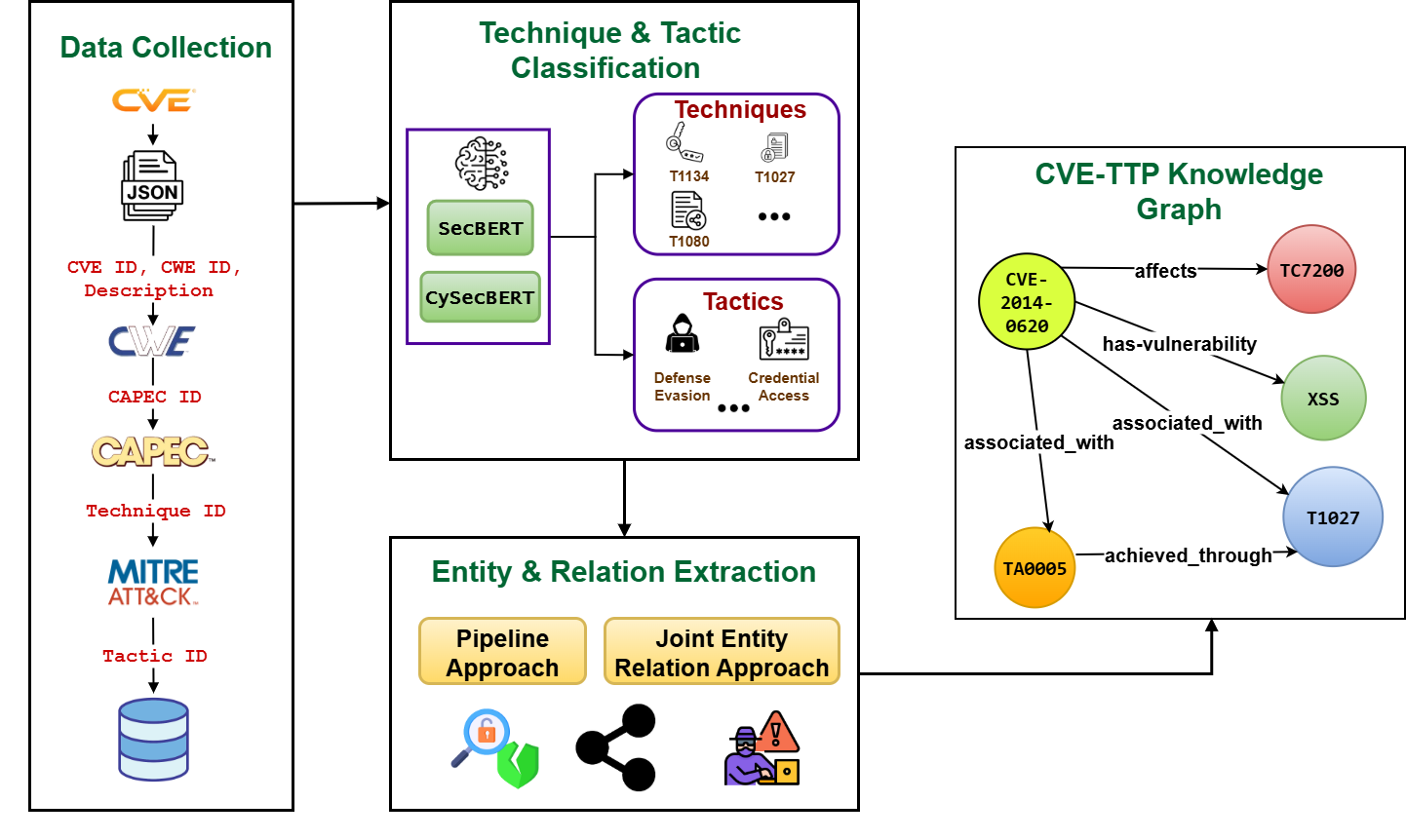}
    \caption{Architecture of the CVE-TTP Knowledge Graph. The construction process consists of four stages: \textit{(i)} data collection, where vulnerability information is gathered from multiple sources; \textit{(ii)} classification, where CVEs are mapped to ATT\&CK techniques and tactics using a prediction model; \textit{(iii)} entity and relation extraction, which identifies structured information; and \textit{(iv)} knowledge graph construction, where the extracted triples are organized into the graph.}
    \label{fig:cve-ttp-kg-architecture}
\end{figure}

This section outlines the architecture of the proposed CVE-TTP KG system, as illustrated in  Figure \ref{fig:cve-ttp-kg-architecture}. The ultimate goal of our system is to construct a comprehensive KG by transforming extracted CVE descriptions into a curated, semantically enriched format designed to capture adversarial behavior patterns. The workflow is structured into four primary modules: \textit{(i)} Data Collection, \textit{(ii)} Linking CVE descriptions with techniques and tactics, \textit{(iii)} Entity and Relation Extraction, \textit{(iv)} KG Generation. Each module is responsible for a specific task within the pipeline.  
   
\subsection{Data Collection}
\label{sec:data_collection}
\begin{figure}[h]
    \centering
    \includegraphics[width=\textwidth]{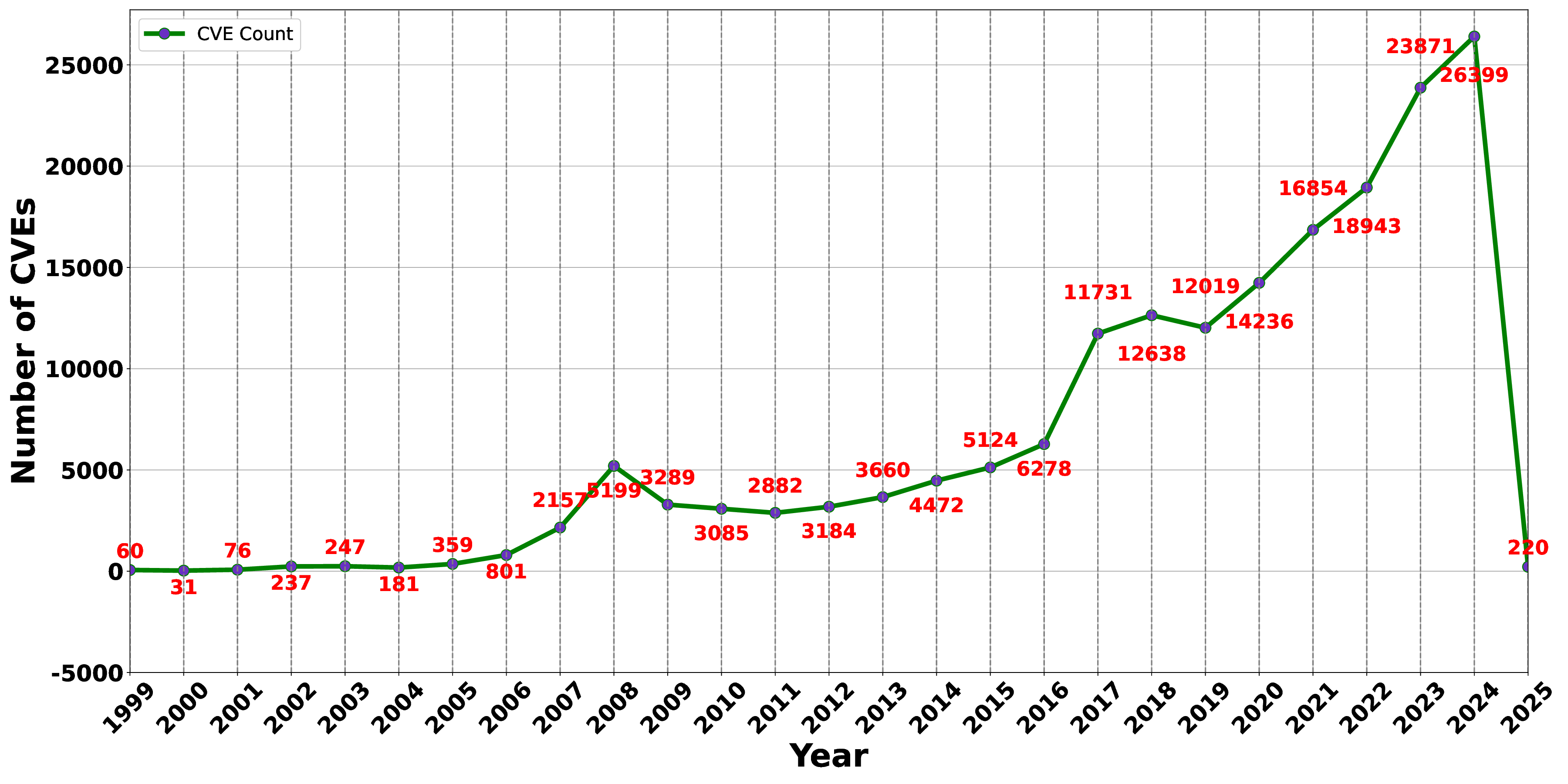} 
    \caption{Year-wise distribution of CVEs. The plot depicts the annual distribution of collected vulnerabilities, showing the number of CVE entries extracted for each year.}
    \label{fig:yearly-cve}
\end{figure}

Initially, we constructed a CVE dataset covering the period 1999--2025 using records from the NVD and the MITRE CVE List V5. NVD provided structured CVE records from 2002 onward, while CVE List V5\footnote{\url{https://github.com/CVEProject/cvelistV5}} was used to supplement the years 1999--2001 and ensure historical completeness. After data collection, a total of 276,289 CVE records were obtained. Figure~\ref{fig:yearly-cve} presents the year-wise distribution of collected CVEs, showing a steady increase in reported vulnerabilities from 1999 to 2024.


To enrich CVEs with adversarial behavior, we integrated information from CWE, CAPEC, and the MITRE ATT\&CK framework using a multi-stage mapping process. First, CWE identifiers were extracted from NVD records. For example, \texttt{CVE-2021-22909}\footnote{\url{https://www.cve.org/CVERecord?id=CVE-2021-22909}}, a Man-in-the-Middle vulnerability in EdgeMAX EdgeRouter firmware updates, is associated with \texttt{CWE-300} (Channel Accessible by Non-Endpoint)\footnote{\url{https://cwe.mitre.org/data/definitions/300.html}}. This CWE maps to CAPEC attack patterns such as \texttt{CAPEC-94} (Adversary in the Middle) and \texttt{CAPEC-57}\footnote{\url{https://capec.mitre.org/data/definitions/57.html}}, which further align with ATT\&CK Technique \texttt{T1040} (Network Sniffing)\footnote{\url{https://attack.mitre.org/techniques/T1040/}}. The technique is linked to the \texttt{Credential Access} and \texttt{Discovery} tactics.

This enrichment pipeline (CVE~$\rightarrow$~CWE~$\rightarrow$~CAPEC~$\rightarrow$~Technique~$\rightarrow$~Tactic) was used to generate ground-truth labels for classification. CVEs without valid CWE mappings were excluded. Of the 276,289 collected CVEs, 178,233 were successfully enriched, resulting in a technique dataset covering 95 ATT\&CK techniques and a tactic dataset spanning all 14 ATT\&CK tactic categories.
This approach forms a solid foundation but suffers from inherent limitations due to the sparse and incomplete mappings available across public resources, particularly from CWE to CAPEC to ATT\&CK techniques. As a result, this method enriches only a subset of CVEs with behavioral context. To overcome this limitation, we developed an automated classification model that predicts ATT\&CK tactics and techniques directly from CVE textual descriptions.
\subsection{Linking CVE description with Techniques and Tactics}
\label{subsec:linking_ttps}
In this phase, we developed an automated tool based on transformer-based models to predict the techniques and tactics associated with a given vulnerability description. This model leverages the CVE dataset constructed in the previous phase (Session \ref{sec:data_collection}). Specifically, each input instance to the model was formulated by combining key contextual information, including the CVE identifier, its textual description, and the associated CWE identifier. The input format was structured as: \textit{``The vulnerability identified as CVE ID: 
\textless CVE ID \textgreater is described as: \textless Description \textgreater. It is associated with CWE ID:  \textless Problem Type (CWE) \textgreater.''}

This dataset was then used to train multi-label classification models for predicting adversarial behaviours.
To perform technique and tactic classification, we employed pretrained BERT-based language models that have been specifically adapted to the cybersecurity domain. \textbf{SecBERT} is a BERT model pretrained on cybersecurity corpora such as APTnotes, STUCCO-Data, CASIE, and SecureNLP, and we used the \texttt{jackaduma/SecBERT} version from Hugging Face. \textbf{CySecBERT}\cite{bayer2024cysecbert} is a cybersecurity-adapted BERT model pretrained on blogs, arXiv, NVD, and Twitter data, and we used \texttt{markusbayer/CySecBERT} for fine-tuning on our dataset.

\subsection{Entity and Relation Extraction}
\label{subsec:entity_relation_extraction}
In this phase, we identify and extract the cybersecurity-specific entities and their semantic relationships, particularly those related to software vulnerabilities. We defined the entities and relationships by reviewing prior works \cite{kuppa2021linking,sun2023multi,shi2024uncovering} on software vulnerability entity and relation extraction. Furthermore, to ensure consistency with established standards in CTI, the entity-relation schema is aligned with the structural specifications of the STIX 2.0 framework~\footnote{\url{https://oasis-open.github.io/cti-documentation/stix/intro.html}}. The identified entities include vulnerability identifiers such as CVE IDs and CWE IDs, Product Name, Product Version, Vendor Name, Vulnerability Type, and Impact. To align with adversarial behavior modeling, the schema also includes Tactic and Technique entities. 

Based on these entities, several key relationships are defined to capture meaningful connections within the KG. The \textit{affects} relationship links a vulnerability to a specific product or hardware it impacts (e.g., CVE-2021-29529 affects TensorFlow). The \textit{has\_version} relationship associates a product with its specific version (e.g., TensorFlow has version TensorFlow 2.4.1), while \textit{has\_vendor} identifies the organization responsible for the product (e.g., TensorFlow has vendor Google). The \textit{has\_weakness} relationship connects a vulnerability or product to a corresponding CWE category (e.g., CVE-2021-29529 has weakness CWE-131). The \textit{has\_impact} relationship represents the effect of a vulnerability on security properties such as confidentiality, integrity, or availability. Furthermore, the \textit{associated\_with} relationship captures links between vulnerabilities and adversarial techniques or between weaknesses and their related CWE categories (e.g., CVE-2021-29529 associated with T1021, and Buffer Overflow associated with CWE-120). The \textit{related\_to} relationship is used to denote connections between similar or linked vulnerabilities (e.g., CVE-2009-2879 related to CVE-2009-2876). Finally, the \textit{achieved\_through} relationship links a tactic to the technique used to accomplish it (e.g., TA0002 achieved through T1204).

After defining the target entity categories and relation types, we manually annotated and constructed a gold-standard dataset for entity and relation extraction using \textit{Label Studio}\footnote{\url{https://labelstud.io/}}. Due to the time-intensive nature of manual annotation, we selected a representative sample of $1,080$ descriptions from the $178,233$ CVE records, based on a $95$\% confidence level and a $3$\% margin of error. 
To perform entity and relation extraction, we experimented with both pipeline-based and joint extraction approaches. The pipeline method first identifies entities, which are then passed to a separate relation extraction model to determine the semantic links between them. In contrast, the joint extraction approach simultaneously detects entities and their relations, constructing relational triples in the form 
\textit{\textless subject, relation, object \textgreater}.

For the pipeline approach, entity recognition is formulated as a BIO tagging task, where each token is labeled as B (beginning), I (inside), or O (outside) of an entity. For example, in ``CVE-2023-1234 affects Microsoft Windows systems'', ``CVE-2023-1234'' is tagged as B-CVE\_ID, while ``Microsoft Windows'' is tagged as B-PRODUCT and I-PRODUCT. For the relation extraction stage, the two target entities in each instance ($e_1$, $e_2$) were highlighted in the input sentence by surrounding $e_1$ with the special marker ``\$\!'' and $e_2$ with ``\#\!''.

For the joint extraction model, the input is represented in the SciERC format, 
which consists of: {\em(i)} \texttt{tokens} --- tokenized CVE descriptions; {\em(ii)} \texttt{entities} --- labeled spans with type and index positions; and {\em(iii)} \texttt{relations} --- semantic links between entity pairs. 
The resulting labeled dataset consists of $1,080$ CVE descriptions, encompassing $24,820$ entities and $43,608$ relations for training and evaluating entity and relation extraction models.

\subsubsection{Pipeline-based Entity and Relation Extraction:}
\begin{figure}
    \centering
    \includegraphics[scale=0.23]{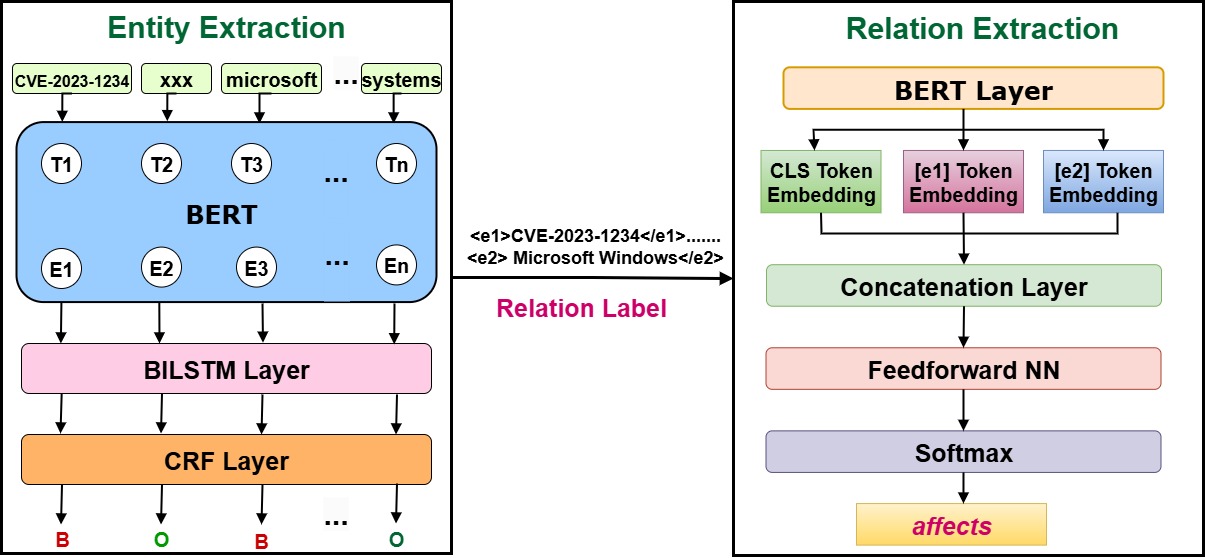}
    \caption{Workflow of pipeline-based entity and relation extraction}
    \label{fig:cve_pipeline}
\end{figure}

In this approach, entity and relation extraction are performed sequentially through a two-stage: {\em(i)} Cybersecurity Entity Recognition (CER) and 
{\em(ii)} Relation Extraction (RE).
The CER stage is designed to identify cybersecurity-related entities from vulnerability (CVE) descriptions, while the RE stage focuses on uncovering semantic associations between these entities. The workflow of the pipeline-based entity and relation extraction is depicted in Figure~\ref{fig:cve_pipeline}.
\paragraph{Cybersecurity Entity Recognition:}  
In the CER stage, we adopt a BERT--BiLSTM--CRF architecture, which has been widely recognized for its effectiveness in domain-specific sequence labeling tasks~\cite{arikkat2024ostis,dasgupta2020comparative,wang2022aptner,wang2022cyber}.  BERT (\texttt{bert-base-uncased}) is used as the encoder to generate contextual token embeddings that capture semantic and technical information. These embeddings pass through a BiLSTM layer to model sequence context in both directions, followed by a CRF layer that ensures consistent label sequences and improves recognition of ambiguous or multi-token cybersecurity entities. 

\paragraph{Relation Extraction:}  
In the RE stage, we identify the relations between entities. 
We adopted a transformer-based \texttt{bert-base-uncased} classifier, motivated by the demonstrated dominance of BERT-based methods in achieving state-of-the-art performance for relation extraction across diverse domains~\cite{diaz2025survey}. 
To guide the model’s attention toward the target entities $e_1$ and $e_2$, we enclose them with special markers (``\$'' for $e_1$ and ``\#'' for $e_2$) in the input sentence. 
The annotated text is then encoded by BERT into contextualized hidden representations. Entity span embeddings are derived via pooling, concatenated, and fed into a fully connected layer with a softmax activation function to predict the relation type.  
\subsubsection{Joint Entity and Relation Extraction:}
\begin{figure}[htbp]
    \centering
    \includegraphics[width=\textwidth]{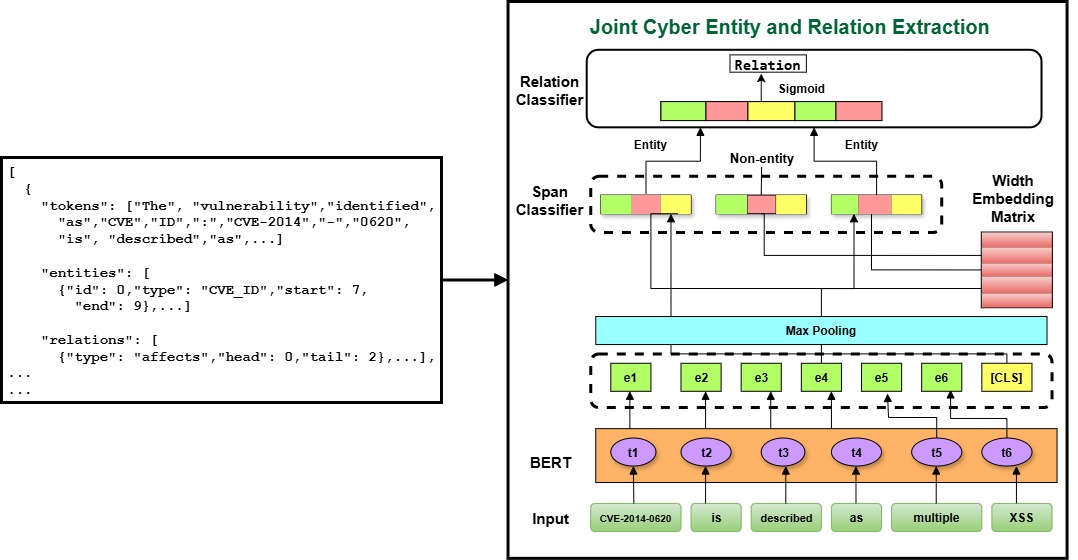}
    \caption{Workflow for joint entity and relation extraction}
    \label{fig:joint_ner_re}
\end{figure}

Traditional pipeline architectures for entity and relation extraction suffer from error propagation and fail to capture the mutual dependencies between entities and their relations \cite{guo2021cyberrel}. To address these limitations, we investigated a joint extraction framework that performs both tasks jointly. Specifically, we utilize a span-based transformer framework \cite{eberts2019span} that processes CVE vulnerability descriptions to extract entity-relation triples in the format $(\texttt{subject}, \texttt{relation}, \texttt{object})$. The overall workflow of the joint entity and relation extraction process is illustrated in Figure~\ref{fig:joint_ner_re}.

The core architecture of the joint entity and relation extraction model is a pre-trained CySecBERT. Instead of token-level labeling, the model operates on spans (contiguous sequences of tokens), treating each possible span as a candidate for entity classification. This method avoids traditional decoding strategies like CRFs or BIO tagging. The joint entity model operates in three stages: \textit{(i)} span classification, \textit{(ii)} span filtering, and \textit{(iii)} relation classification.
\paragraph{Entity Identification via Span Classification:}
For each sentence, the model generates contextual embeddings for all tokens, including a special \texttt{[CLS]}(classifier) token that represents the overall context of the sentence. Candidate spans are constructed from consecutive token sequences, each of which is encoded using max pooling over its token embeddings.
To enrich the span representation, two additional vectors are concatenated: \textit{(i)} a width embedding that captures the length of the span, and \textit{(ii)} the \texttt{[CLS]} embedding to integrate sentence-level information. This concatenated vector is passed through a softmax classifier to predict an entity type or a non-entity class.
\paragraph{Span Filtering:}
Spans predicted as non-entities are filtered out. To reduce computational overhead, we discard spans exceeding a predefined maximum length before classification. The resulting set comprises only those spans that are likely to represent meaningful entities.
\paragraph{Relation Classification:}
Each pair of retained entity spans is evaluated to determine whether a semantic relation exists between them. For each ordered pair $(s_1, s_2)$, a composite vector is formed by concatenating: \textit{(i)} the span embeddings of $s_1$ and $s_2$, \textit{(ii)} their respective width embeddings, and \textit{(iii)} a context embedding derived from the token embeddings found between the two spans.

If the span pair is adjacent or overlapped, the context vector is replaced by a zero vector. This combined representation is fed into a sigmoid classifier to determine the presence of one or more relations from a predefined set. Since relations can be directional, both $(s_1, s_2)$ and $(s_2, s_1)$ are evaluated.
\subsection{CVE-TTP Knowledge Graph Construction}
\label{subsec:cve_kg}
A Cybersecurity KG (CKG) provides a structured and semantically rich representation of cybersecurity-related concepts. In this framework, key entities such as vulnerabilities, threat actors, exploits, attack tactics, techniques, etc. are modeled as nodes, while the semantic relationships among them are represented as directed edges. This KG provides a unified structure that bridges low-level technical vulnerabilities with high-level adversarial strategies.
The construction of the CVE-TTP KG involves the following key stages:

\begin{itemize}
    \item \textbf{Entity and Relation Extraction:} Entities and their semantic relationships are extracted from CVE vulnerability descriptions using the extraction methods described in the Section \ref{subsec:entity_relation_extraction}.
    
    \item \textbf{Graph Generation using Neo4j:} The extracted entity-relation triples are used to populate a CVE-TTP graph using Neo4j \footnote{https://neo4j.com/}. Neo4j, a native graph database, enables efficient storage, traversal, and querying of the graph structure using the Cypher query language. This facilitates downstream analytics such as subgraph exploration, threat attribution, vulnerability prioritization, and adversary behavior mapping. Each unique entity is represented as a node with associated metadata (e.g., type, name, identifier), and each relation is stored as a directed edge connecting two nodes, labeled with the corresponding relation type. 
\end{itemize}

\section{Experiment \& Analysis}
\label{sec:experiment_results}
In this section, we detail the experimental setup, followed by a discussion of the results and their analysis.
\subsection{Experimental Setup}
The experimentation was conducted using Python on a local machine powered by an Apple M3 chipset. The system featured an 8-core integrated GPU and leveraged Apple’s Metal 3 API for hardware acceleration. We employed libraries, including PyTorch and Hugging Face Transformers for model training and inference, scikit-learn for performance evaluation, matplotlib for result visualization, and the Neo4j Python driver for constructing the KG.

To evaluate the performance of the proposed model across different tasks, we employed metrics appropriate to each problem. Since the classification of tactics and techniques involves predicting multiple labels per instance, we adopted evaluation metrics commonly used in multi-label classification. These include Hamming Loss and Jaccard Similarity, along with standard metrics including Precision, Recall, F1-Score, and Accuracy \cite{riera2022new}. For entity recognition and relation extraction tasks, we evaluated model performance using Precision, Recall, F1-Score, and Accuracy to assess the quality and correctness of the extracted entities and their semantic relations.
\subsection{Performance on Technique and Tactic Classification}
This section presents the results of the proposed model for technique and tactic classification from vulnerability descriptions. For the experimentation, we have used the labeled dataset generated through the process outlined in Section~\ref{sec:data_collection}. 
The technique classification dataset contains 178,233 vulnerability descriptions annotated with 95 unique techniques, while the tactic classification dataset includes the same 178,233 descriptions labeled with 14 unique tactics. To ensure the reliability of supervised learning, we excluded labels with insufficient samples. Specifically, techniques and tactics with fewer than $50$ instances were discarded. After this filtering, the final dataset used for model training includes $80$ techniques and $13$ tactics.
The dataset was split into 80:10:10 for training, validation, and testing. For technique and tactic prediction, we used pre-trained cybersecurity models, SecBERT and CySecBERT, fine-tuned on our dataset with five random seeds for stability. Experiments used the AdamW optimizer with a learning rate of $1 \times 10^{-5}$ for up to five epochs. Batch sizes were set to 8 for \texttt{CySecBERT} and 16 for \texttt{SecBERT} based on GPU constraints. Binary Cross-Entropy with Logits was used to handle the multi-label classification task.


The performance on the test set for both technique and tactic classification is summarized in Table \ref{tab:classification_results}. 
Overall, CySecBERT consistently outperforms SecBERT in both tasks.
In the technique classification task, CySecBERT achieves a higher Jaccard score (0.9659) and F1-macro (0.8771). 
Similarly, Hamming loss is lower for CySecBERT (0.0071). 
For the tactic classification task, both models achieve even higher performance across all metrics, with CySecBERT again outperforming SecBERT. Specifically, CySecBERT reaches a Jaccard score of $0.9858$, F1-micro of $0.9899$, and F1-macro of $0.9616$, reflecting generalization across all tactic labels. 

\begin{table}[htbp]
\fontsize{6}{7}\selectfont
\caption{Average performance of SecBERT and CySecBERT models for technique and tactic classification, averaged over five different random seed runs.}
\centering
\renewcommand{\arraystretch}{1}
\setlength{\tabcolsep}{8pt}
\begin{tabular}{|l|cc|cc|}
\hline
 & \multicolumn{2}{c|}{\textbf{Technique Classification}} & \multicolumn{2}{c|}{\textbf{Tactic Classification}} \\
\hline
\textbf{Metric (Avg.)} & \textbf{SecBERT} & \textbf{CySecBERT} & \textbf{SecBERT} & \textbf{CySecBERT} \\
\hline
Jaccard Score          & 0.9584 & 0.9659 & 0.9829 & 0.9858 \\
Hamming Loss           & 0.0085 & 0.0071 & 0.0084 & 0.0071 \\
F1 Micro               & 0.9582 & 0.9657 & 0.9879 & 0.9899 \\
F1 Macro               & 0.8361 & 0.8771 & 0.9530 & 0.9616 \\
F1 Weighted            & 0.9574 & 0.9651 & 0.9877 & 0.9898 \\
Precision Micro        & 0.9749 & 0.9797 & 0.9920 & 0.9929 \\
Precision Macro        & 0.9199 & 0.9334 & 0.9823 & 0.9849 \\
Precision Weighted     & 0.9744 & 0.9794 & 0.9919 & 0.9928 \\
Recall Micro           & 0.9421 & 0.9517 & 0.9839 & 0.9869 \\
Recall Macro           & 0.7927 & 0.8381 & 0.9281 & 0.9413 \\
Recall Weighted        & 0.9421 & 0.9517 & 0.9839 & 0.9869 \\
Accuracy               & 0.9159 & 0.9379 & 0.9461 & 0.9552 \\
\hline
\end{tabular}
\label{tab:classification_results}
\end{table}
\subsection{Performance on Entity and Relation Extraction}
We utilized the labeled dataset constructed through the steps outlined in Section~\ref{subsec:entity_relation_extraction}. The dataset comprises $1,080$ CVE vulnerability descriptions with $24,820$ entities and $43,608$ relations. 
The dataset was divided into training, validation, and test sets with an 80:10:10 proportion.
\subsubsection{Pipeline Approach:}
In the first stage of our pipeline approach, we extracted the entities using BERT--BiLSTM--CRF architecture. The \texttt{bert-base-cased} encoder first generates contextual embeddings, which are passed through a BiLSTM with a hidden size of 256, a fully connected layer, and a CRF layer for tag prediction. Training was carried out for 10 epochs with the AdamW optimizer, a maximum sequence length of 256 tokens, a learning rate of $2\times 10^{-5}$, a batch size of 16, and a dropout rate of 0.1. 
The proposed CER module achieved an F1 micro score of $0.98$, an F1 weighted score of $0.99$, and an F1 macro score of $0.86$ on the test set. The second stage identifies semantic relations between extracted entity pairs using \texttt{bert-base-uncased} with a maximum sequence length of 512 tokens. The model was trained for one epoch using Adam, a learning rate of $1\times10^{-5}$, batch size 16, and dropout $0.2$ before each dense layer.
The relation extraction (RE) system in the pipeline approach achieved an F1-score of $0.99$, with a precision of $0.9961$ and a recall of $0.9960$. The confusion matrices for the CER and RE stages are presented in Appendix Figure~\ref{fig:cm_cer} and Figure~\ref{fig:cm_re}.
\subsubsection{Joint Entity and Relation Extraction Approach:}
For joint entity and relation extraction, we used a span-based Entity–Relation Transformer~\cite{eberts2019span} with CySecBERT. The model was trained for 25 epochs with a batch size of 2, using a learning rate of $2e-5$, linear warmup of $0.1$, and weight decay of $0.01$. Gradient clipping ($1.0$) and dropout ($0.2$) were applied, with a maximum span size of 10. 
To balance samples, we used 300 negative entities and 200 negative relations per batch, limiting candidate span pairs to 750 per document and applying a relation threshold of $0.5$. A span size embedding of 25 was included, and training was parallelized using four processes. Early stopping was based on validation performance.
For entity extraction, the joint model achieved a micro-F1 score of $95.86$ and a macro-F1 score of $88.20$. 
 
Furthermore, we evaluated the relation extraction performance of the joint entity–relation model under two settings: the NEC (Named Entity Classification) constraint and the non-NEC (span-only) condition. In the NEC setting, evaluation is stringent, requiring that each predicted relation instance satisfy three criteria: (i) both entity spans are correctly identified, (ii) the entity types are accurately classified, and (iii) the predicted relation label matches the gold standard. In contrast, the non-NEC setting (also referred to as a ``relaxed" evaluation) requires only that the entity spans and the relation label are correct, without enforcing the correctness of entity type assignments. Under these settings, the joint model achieved a micro-F1 of $77.81$ and a macro-F1 of $78.70$ in the non-NEC scenario, and a micro-F1 of $77.77$ with a macro-F1 of $78.69$ in the NEC scenario. The detailed class-wise performance of entities and relations is presented as a confusion matrix in Appendix Figure~\ref{fig:cm_joint_ent} and Figure~\ref{fig:cm_joint_rel}.
\subsection{CVE-TTP Knowledge Graph Generation}
The construction of the CVE-TTP KG transforms outputs from entity and relation extraction models into structured triples, which are then imported into Neo4j for visualization and analysis. 
For the joint extraction model, predictions are stored in JSON format, containing tokens, identified entities, relations, and their spans. These outputs are programmatically converted into $\langle$\textit{head, relation, tail}$\rangle$ triples to semantically represent connections between entities. In contrast, the pipeline model produces triples composed of the \textit{head}, \textit{relation}, and \textit{tail} without span information, since entity and relation predictions are generated in separate sequential steps.  For the KG generation task, we employed the same CVE descriptions to assess the performance of both the pipeline and joint models. Specifically, we used 50 vulnerability descriptions that were not included in the training phase of any model to generate triples. For demonstration purposes, we illustrate the process using the CVE-2020-0617 vulnerability:

\begin{quote}
\scriptsize
    The vulnerability identified as CVE ID: \texttt{CVE-2020-0617} is described as: A denial of service vulnerability exists when Microsoft Hyper-V Virtual PCI on a host server fails to properly validate input from a privileged user on a guest operating system, also known as ``Hyper-V Denial of Service Vulnerability''. It is associated with \texttt{CWE ID: CWE-20}. For organizations monitoring cyber threat activities, this vulnerability aligns with Tactic ID(s): \texttt{TA0003, TA0004, TA0005, TA0006} and Technique ID(s): \texttt{T1027, T1036.001, T1539, T1553.002, T1562.003, T1574.006, T1574.007}, representing a critical threat to security measures and warranting immediate attention.
\end{quote}

\begin{figure}[!h]
\centering
\includegraphics[width=0.8\textwidth]{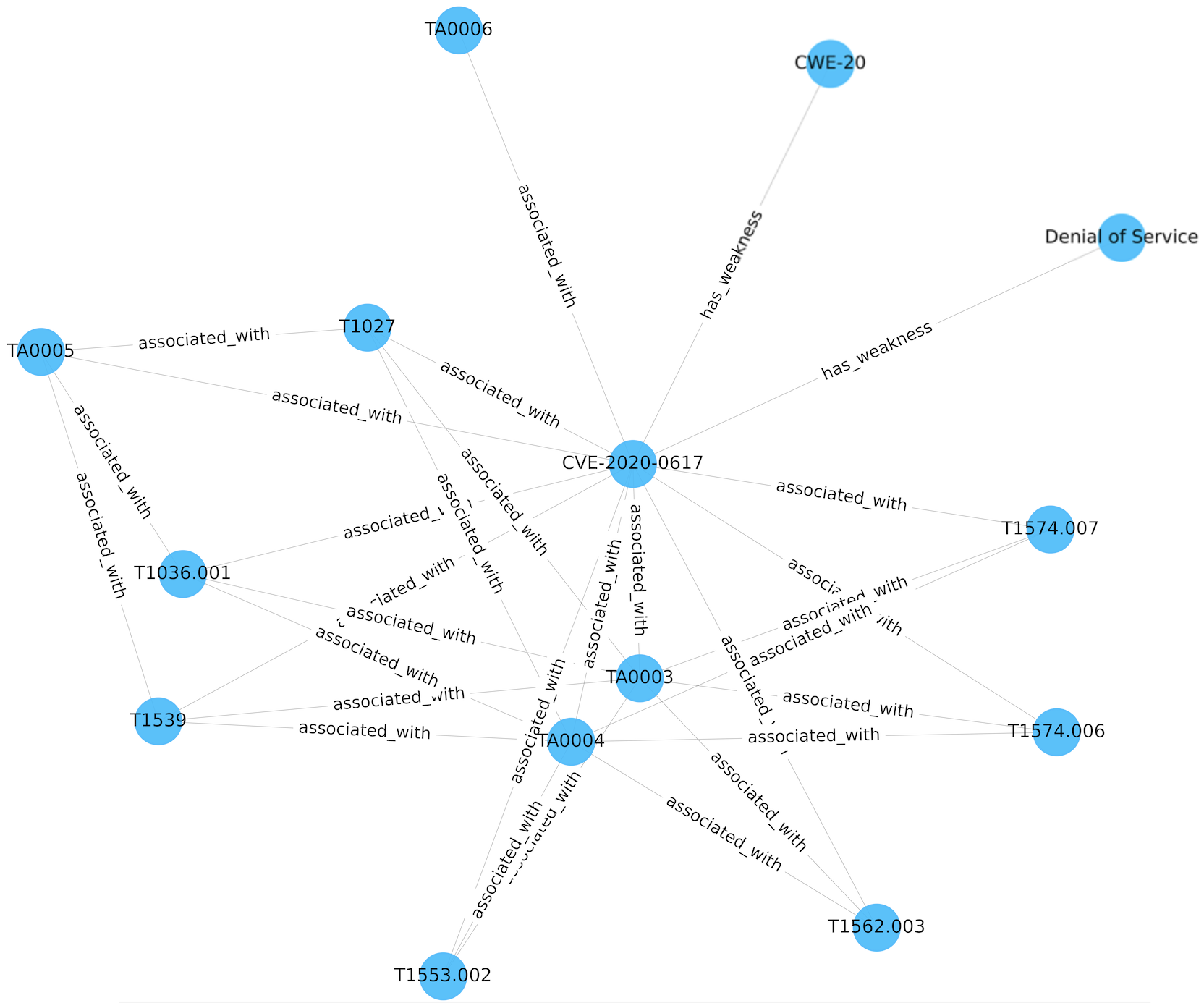}
\caption{CVE-TTP KG generated by the pipeline model, illustrating entity associations alongside semantic inconsistencies such as generic tactic-technique relations.}
\label{fig:cve_ttp_kg_pipeline}
\end{figure}

\paragraph{Pipeline Model KG Generation:}

The $\langle$\textit{head, relation, tail}$\rangle$ triples extracted using CER and relation extraction are imported into Neo4j for KG construction and visualization. For the CVE-2020-0617 vulnerability, the pipeline model's KG includes (Figure ~\ref{fig:cve_ttp_kg_pipeline}) key entities such as \texttt{CVE-2020-0617} (vulnerability identifier), \texttt{CWE-20} (weakness classification), multiple tactic nodes (\texttt{TA0005}, \texttt{TA0003}, \texttt{TA0004}, \texttt{TA0006}), and technique nodes (\texttt{T1027}, \texttt{T1036.001}, \texttt{T1539}, \texttt{T1553.002}, \texttt{T1562.003}, \texttt{T1574.006}, \texttt{T1574.007}). The relation \texttt{associated\_with} predominates, linking tactics and techniques to the central CVE node. However, this model often replaces the semantically accurate \texttt{achieved\_through} relation, which specifies how tactics are operationalized through techniques, with the more generic \texttt{associated\_with} relation. This substitution reduces the granularity and semantic precision in mapping procedural cybersecurity workflows.
Furthermore, while the pipeline model correctly predicts the \texttt{has\_weakness} relation between CVE and CWE entities, it misclassifies ``Denial of Service'' as a weakness rather than properly categorizing it as a vulnerability type. This misclassification adversely affects the accuracy of representing vulnerability impacts, unlike the joint model, which successfully captures this context. 
\begin{figure}[!h]
\centering
\includegraphics[width=0.8\textwidth]{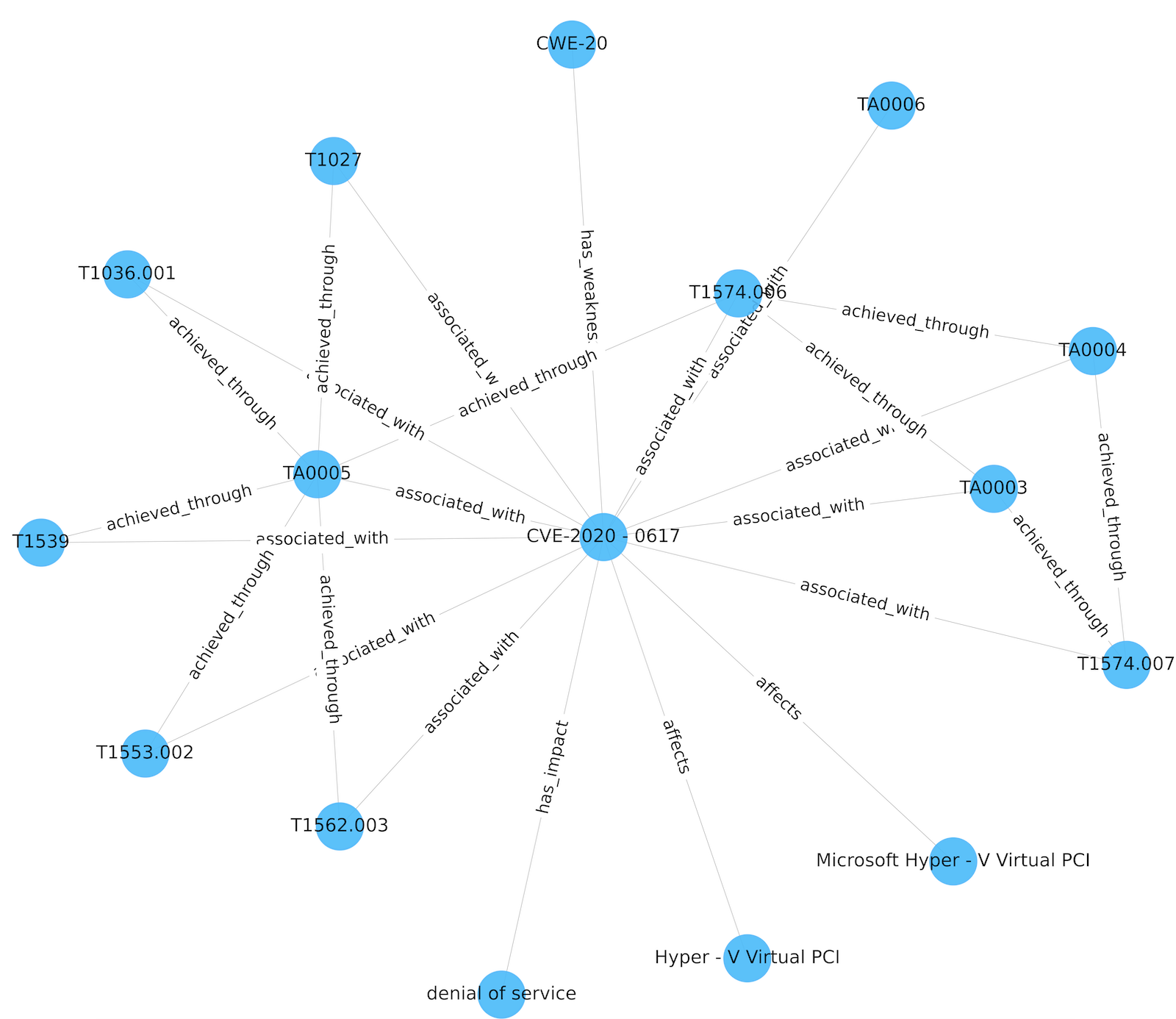}
\caption{CVE-TTP KG generated by the joint model includes CVE, CWE, product, tactic, and technique entities with semantically precise relations.}
\label{fig:cve_ttp_kg_joint}
\end{figure}

\paragraph{Joint Model KG Generation:}
Alternatively, the joint extraction model performs entity recognition and relation extraction concurrently, predicting $\langle$\textit{head, relation, tail}$\rangle$ triples directly from text without intermediate entity pairing. This end-to-end inference reduces error propagation found in pipeline architectures and enhances extraction efficiency. The resulting triples were imported into Neo4j for KG construction.
The resulting KG centers on \texttt{CVE-2020-0617} (Figure \ref{fig:cve_ttp_kg_joint}), representing a denial of service vulnerability in Microsoft Hyper-V Virtual PCI stemming from input validation flaws. This node connects to entities such as \texttt{CWE-20} (weakness category), related products (\texttt{Microsoft Hyper-V Virtual PCI}), and the impact type (\texttt{denial of service}). The graph captures extensive security context through tactic nodes (\texttt{TA0005}, \texttt{TA0003}) and technique nodes (\texttt{T1027}, \texttt{T1539}, \texttt{T1533.002}), reflecting mappings aligned to MITRE adversarial patterns.

Relations extracted include \texttt{associated\_with}, linking CVEs to tactics and techniques; \texttt{achieved\_through}, specifying how tactics are executed via techniques or sub-techniques; \texttt{has\_weakness}, mapping CVEs to weakness classifications; \texttt{affects}, connecting vulnerabilities to impacted products; and \texttt{has\_impact}, representing the consequence of the vulnerability. These relations form a semantically rich and technically precise structure emphasizing attack sequences and vulnerability impacts, albeit with less representation of vendor or product version contexts. Compared to the pipeline KG, the joint model produces a more compact and focused KG, with strong CVE-to-CWE and CVE-to-TTP edges, but fewer explicit tactic-to-technique connections. This outcome highlights a trade-off: the joint model excels in end-to-end robust extraction of core vulnerability and weakness mappings while sacrificing some granularity in operational semantics concerning attack workflows.
While the joint model generally predicts \texttt{None} when uncertain, it also exhibits misclassifications between semantically similar entity types. For example, both models failed to correctly identify the \texttt{Vendor} entity ``Microsoft'' and the associated relation \texttt{has\_vendor}~$\langle$\texttt{Hyper-V Virtual PCI}, \texttt{has\_vendor}, \texttt{Microsoft}$\rangle$. The pipeline model omitted this entity and relation entirely, whereas the joint model incorrectly classified ``Microsoft Hyper-V Virtual PCI'' as a \texttt{Product}. This behavior reflects the joint model's integrated prediction mechanism, which strives to assign the most plausible labels jointly under ambiguous contexts, contrasting with the pipeline model's tendency to omit difficult predictions. Despite these misclassifications, the joint model's unified approach mitigates error propagation between entity and relation extraction subtasks, a common limitation in the pipeline design. Consequently, the joint model demonstrates superior performance and robustness for comprehensive KG generation, especially in contexts with interdependent and overlapping entity-relation semantics.
\section{Conclusion}
\label{sec:conclusion}
The rapid increase in software vulnerabilities challenges cybersecurity teams, especially in linking vulnerabilities to real-world attack behaviors. Although databases like CVE and NVD provide technical details, they lack connections to tactics and techniques, limiting effective threat response. In this work, we propose CVE-TTP KG, a framework that extracts cybersecurity entities and relationships from CVE descriptions to construct a structured KG to improve threat analysis. We collected and mapped data linking CVEs with weaknesses, tactics, and techniques, and created two datasets for multi-label classification and entity–relation extraction. Using CySecBERT, we achieved macro F1-scores of 0.8771 for technique classification and 0.9616 for tactic classification. For extraction, the pipeline approach reached 0.86 (entities) and 0.99 (relations), while the span-based model achieved 0.78. The resulting KG provides a clear representation of vulnerability–attack relationships, supporting better analysis and mitigation. Future work includes integrating Large Language Models to improve extraction and addressing challenges like entity disambiguation and coreference resolution. Resolving ambiguous terms (e.g., ``service'') and references (e.g., ``it'' or ``this issue'') will enhance entity linking and improve the accuracy and usefulness of the KG for threat modeling.

\bibliographystyle{splncs04}
\bibliography{references}

\appendix
\section {Class-wise Analysis of Entity Relation Extraction Approaches}
\subsection{Pipeline Approach}
\vspace{-2em}
\begin{figure}[h]
    \centering
    \includegraphics[scale=0.065]{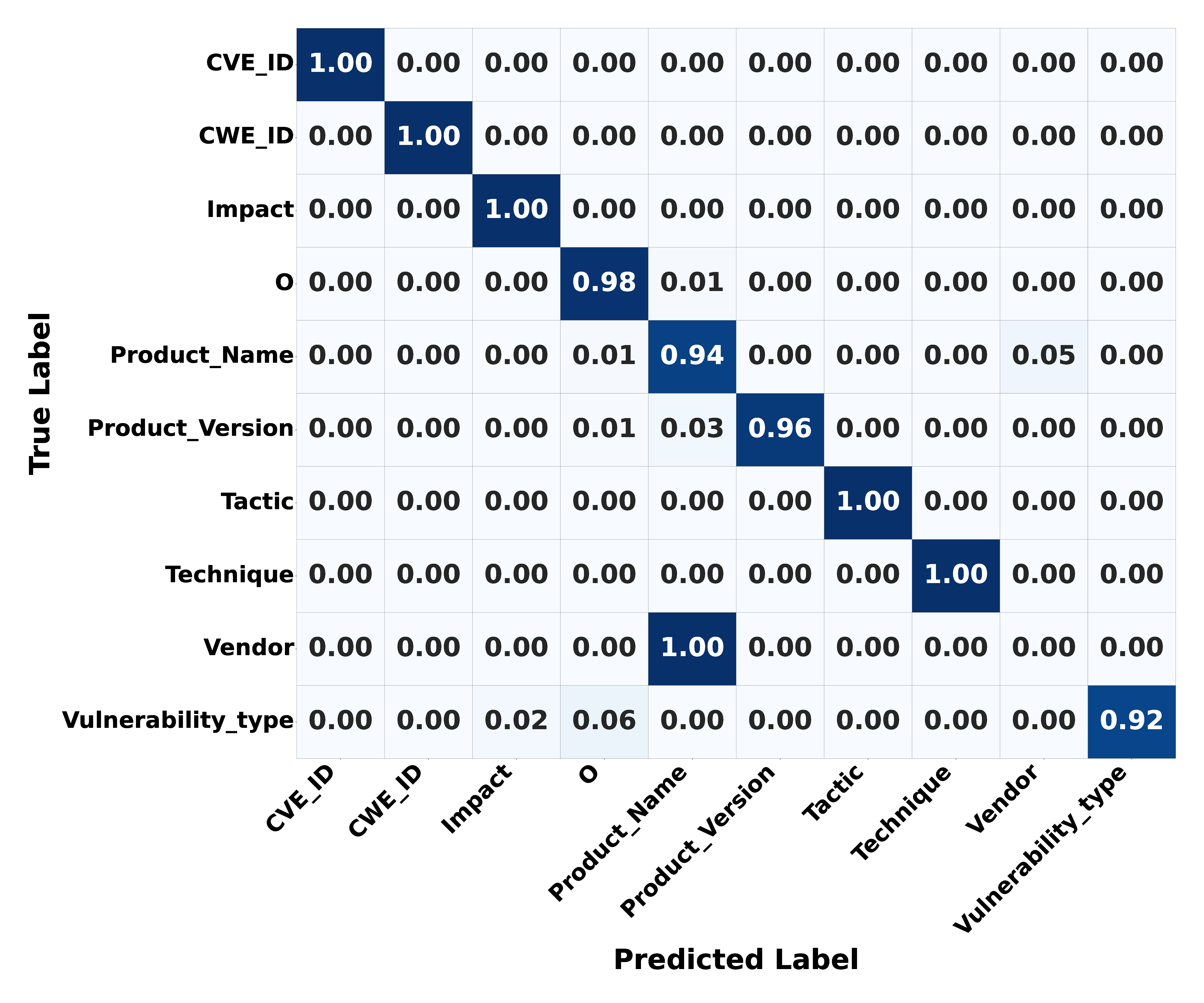}
    \caption{Confusion matrix for the  Entity Recognition stage of the pipeline approach.}
    \label{fig:cm_cer}
\end{figure}

The confusion matrix for the CER phase is presented in Figure~\ref{fig:cm_cer}. From the confusion matrix, we can infer that several entities, including \texttt{CVE\_ID}, \texttt{CWE\_ID}, \texttt{Impact}, \texttt{Tactic}, and \texttt{Technique}, achieved a classification performance of 1.00. 
Conversely, the model's performance decreases for categories exhibiting greater semantic overlap. In particular, \texttt{Vendor} entity is misclassified as \texttt{Product\_Name}. This confusion likely stems from the lexical and contextual similarity between vendor names and product identifiers (e.g., \emph{Microsoft} or \emph{Oracle} can occur as both vendor and product in CVE text).  Overall, the CER module demonstrates strong performance across most entity types. Most misclassifications involve low-frequency classes with overlapping contextual characteristics.

\begin{figure}[h]
    \centering
    \includegraphics[scale=0.065]{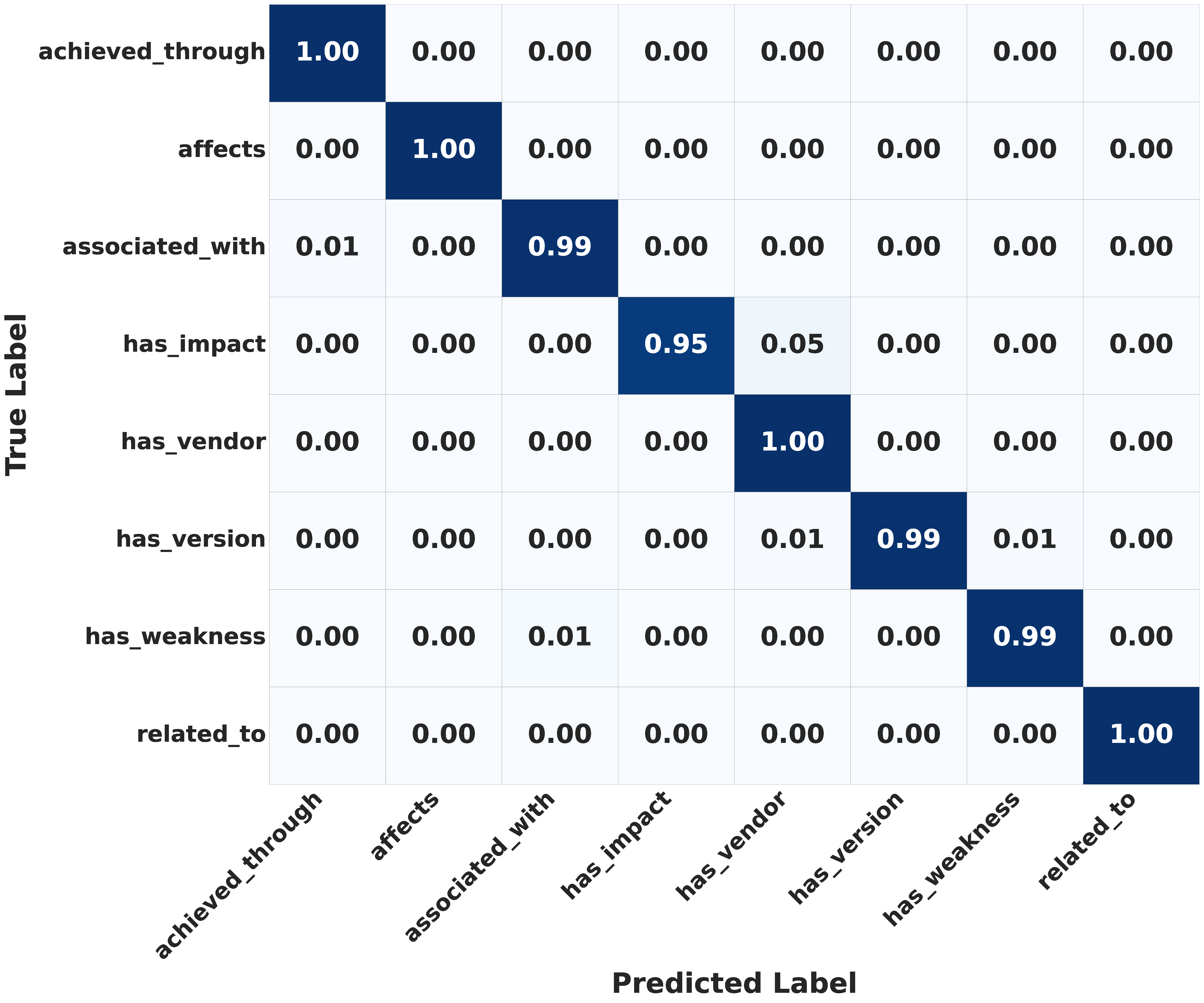}
    \caption{Confusion matrix for the Relation Extraction stage of the pipeline approach.}
    \label{fig:cm_re}
\end{figure}
 The individual relation type performance of the RE task is presented in Figure~\ref{fig:cm_re}. Several relation types, including \texttt{achieved\_through}, \texttt{affects}, \texttt{has\_vendor}, and \texttt{related\_to}, are predicted without error (score of $1.00$). The most pronounced challenge emerged with \texttt{has\_impact}, achieving $0.95$ accuracy, with around $5\%$ of cases re-labelled as \texttt{has\_vendor}. This confusion likely stems from the narrative style of vulnerability descriptions, although the RE component achieved high performance across relation types. 

\subsection{Joint Entity and Relation Extraction Approach}

 \begin{figure}[!h]
    \centering
    \includegraphics[scale = 0.065]{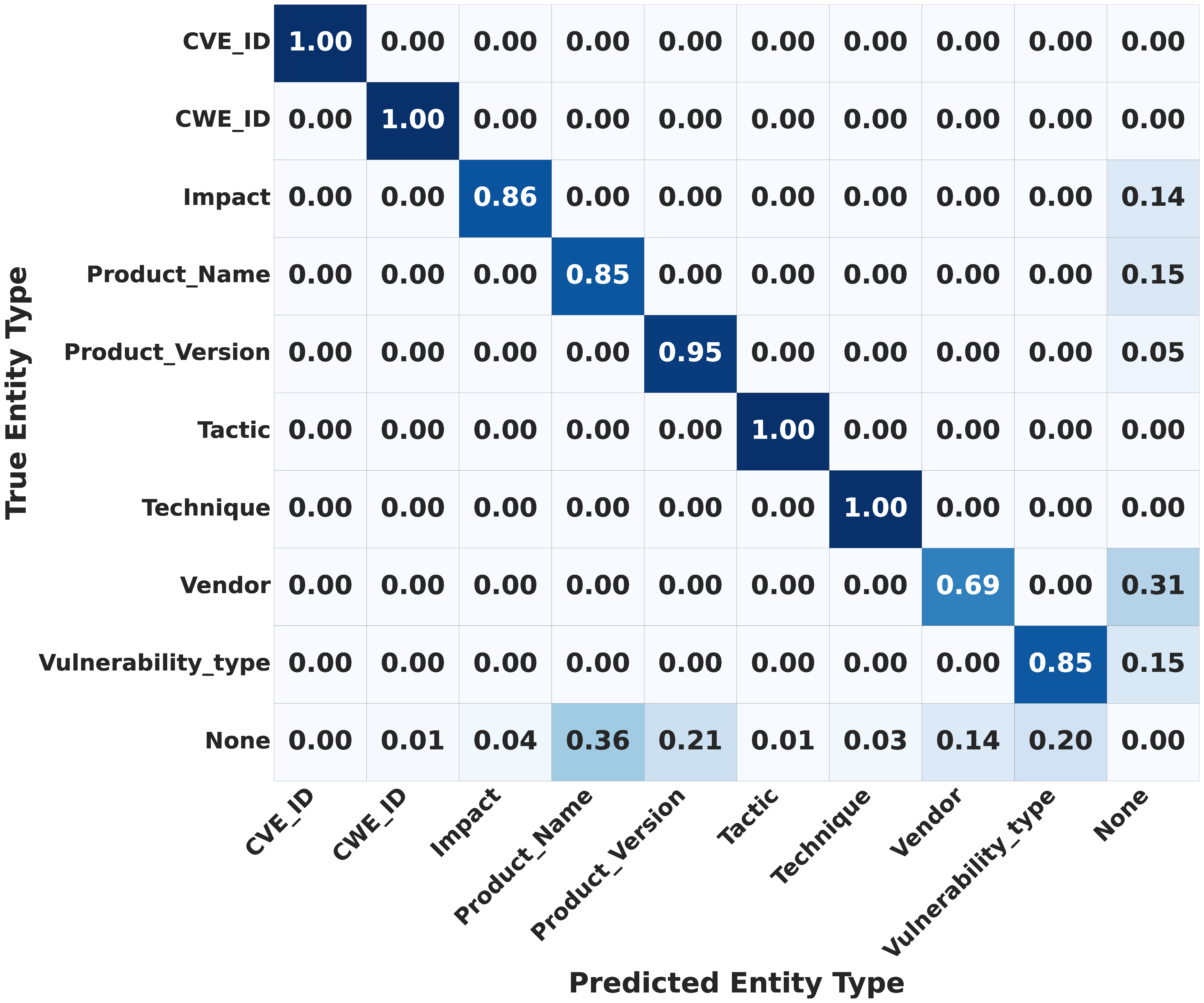}
    \caption{Confusion matrix for entities in the joint model.}
    \label{fig:cm_joint_ent}
\end{figure}
\begin{figure}[!h]
    \centering
    \includegraphics[scale=0.065]{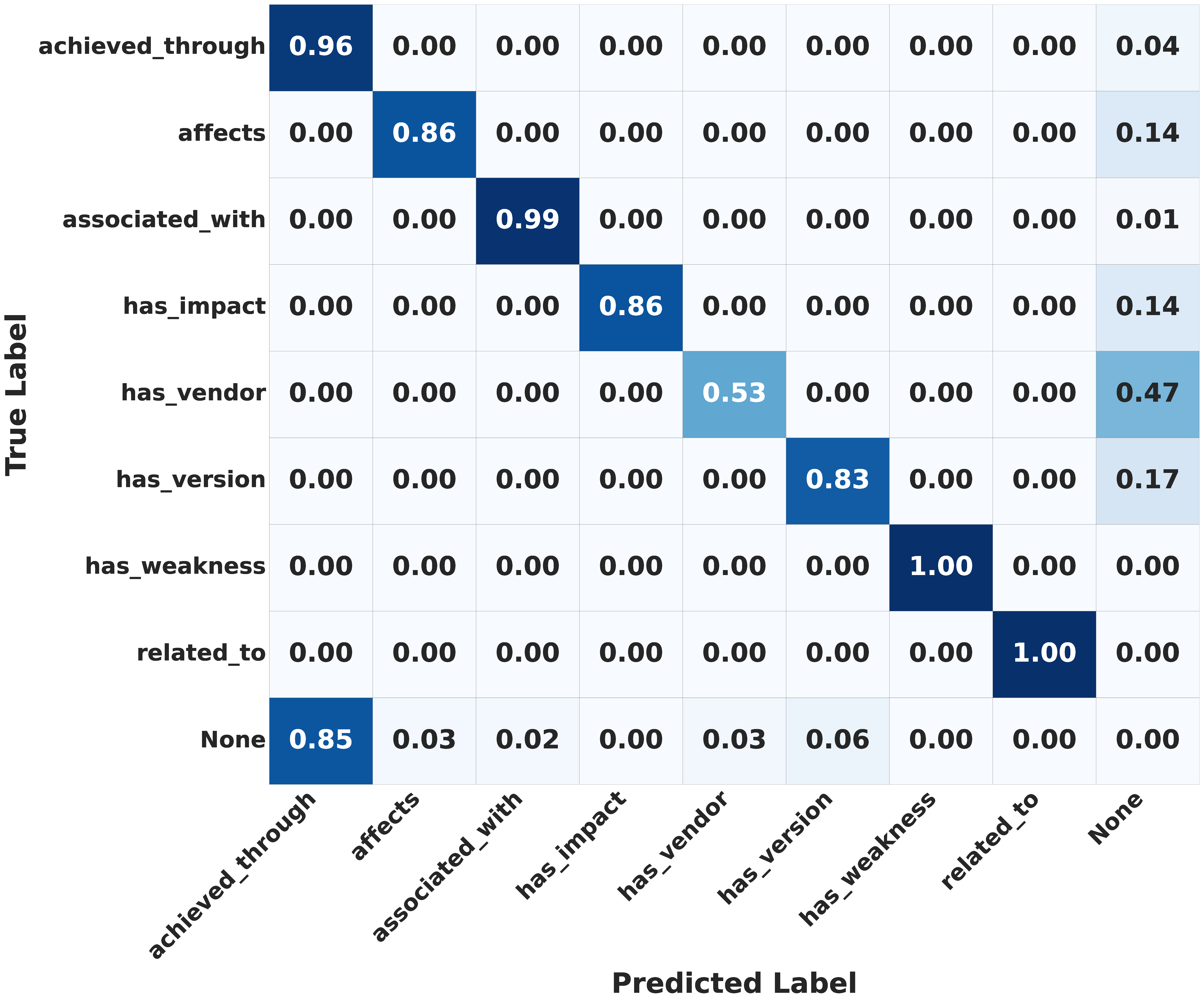}
    \caption{Confusion matrix for relations in the joint model.}
    \label{fig:cm_joint_rel}
\end{figure}
 Figure~\ref{fig:cm_joint_rel} presents the confusion matrix of the entities in the joint model. The results indicate that the model performs exceptionally well on structured identifiers and well-defined categories, including \texttt{CVE\_ID}, \texttt{CWE\_ID}, \texttt{Tactic}, and \texttt{Technique}, with perfect performance of $1.00$, and $0.95$ for \texttt{Product\_Version}. In contrast, the \texttt{Vendor} class exhibits comparatively lower performance ($0.69$), with approximately $31\%$ of instances predicted as \texttt{None}. In the confusion matrices for the joint entity and relation extraction model, the label \texttt{None} signifies that the model predicted the absence of a valid entity or relation.

The confusion matrix shown in Figure~\ref{fig:cm_joint_rel} summarizes the model’s performance on the relations in the joint model. The results indicate strong performance for relation types such as \texttt{has\_weakness} ($1.00$), \texttt{related\_to} ($1.00$), \texttt{associated\_with} ($0.99$), and \texttt{achieved\_through} ($0.96$). For the \texttt{has\_vendor} relation, a reduced performance of $0.53$ was observed. The lower score for \texttt{has\_vendor} is mainly due to weaker \texttt{Vendor} entity predictions, which limit correct relation extraction. In particular, $47\%$ of \texttt{has\_vendor} instances were misclassified as \texttt{None}, showing difficulty in learning this low-frequency relation. 

\end{document}